 \documentclass[final,5p,times,twocolumn]{elsarticle}
\usepackage{amsmath,amssymb,amsfonts}
\usepackage{amsfonts,array, amssymb, bm, bbm,color}
\usepackage[colorlinks=true, pdfstartview=FitV, linkcolor=red, citecolor=cyan, urlcolor=red]{hyperref}
\usepackage{slashed}
\usepackage{bbm}
\usepackage{nicefrac}
\usepackage{dsfont}
\usepackage{fixmath}
\usepackage{mathtools}

\usepackage{amssymb}

\usepackage{amssymb}
\makeatletter
\def\ps@pprintTitle{%
	\let\@oddhead\@empty
	\let\@evenhead\@empty
	\let\@oddfoot\@empty
	\let\@evenfoot\@oddfoot
}
\makeatother

\begin{document}

\begin{frontmatter}



\title{Studying the holographic Fermi surface in the scalar induced anisotropic background }

\author{Sayan Chakrabarti, Debaprasad Maity, and Wadbor Wahlang}

\address[label1]{Department of Physics, Indian Institute of Technology, Guwahati\\
Assam 781039, India}



\begin{abstract}
Holographic properties of a finite density fermion system have been shown to exhibit many interesting behaviours which can be observed in future. In this paper, we study low energy fermion properties in the framework of the holographic Mott-Insulator system. We study the nature of the Fermi surface and its evolution by tuning two types of dipole couplings in the bulk. We further introduce translational symmetry breaking complex scalar field, which is assumed to couple with the holographic fermions. The symmetry breaking background induced by the scalar field is known as Q-lattice. We calculate the fermion spectral function, which captures the low energy behaviour of the system. By tuning the dipole parameters and the non-normalizable component of the scalar field, we observe interesting phenomena such as spectral weight transfer, Fermi surface smearing, which has already been reported in various real condensed matter experiments. 
\end{abstract}


\end{frontmatter}



\section{Introduction}

Holography has been shown to play a very interesting role in various branches of physics, specifically at strong coupling.
The holographic principle \cite{Maldacena:1997re, Witten:1998qj} has been successfully applied as a framework to understand various phenomena: from non-linear hydrodynamics \cite{Bhattacharyya:2008jc}, Fermi liquid behavior \cite{Faulkner:2009wj,Cubrovic:2009ye}, transport phenomena \cite{Herzog:2007ij}, to high temperature superconductors \cite{Hartnoll:2008vx, Hartnoll:2008kx,Horowitz:2010gk} to name a few and it also helps in the understanding the condensed matter systems with the help of classical gravity theories. One of the earlier important applications of the holographic method is in the case of holographic superconductors, and the results were remarkably similar to those obtained by conventional methods. Since then, it was widely used as a tool to study condensed matter systems. Several works have been done related to Fermi surfaces in order to study the emergence of Fermi and non-Fermi liquid behaviours in the fermionic system \cite{Liu:2009dm,Lee:2008xf,Faulkner:2009am,Guarrera:2011my,Faulkner:2011tm}, also this method was used to study the high $T_c$ superconductor, Mott transition and the pseudogap phase \cite{Vanacore:2015poa,Edalati:2010ww,PhysRevD.90.126013} and there of course exists many more applications in the field of nuclear physics. 
Apart from many such known properties of real condensed matter systems, the holographic study, particularly on the strongly coupled fermionic system at finite density has been shown to predict many exotic properties and phases which may be possible to observe in the laboratory in future. Therefore, exploring various seemingly exotic phases of holographic fermionic system, even without being able to be observed in the laboratory in the near future, is being considered as an active research area. Following the same line, in this paper, we study such holographic systems with some specific types of bulk controlling parameters that have been introduced before in the context of holographic Mott-Insulator physics. Our study reveals many interesting features of the Fermi surface, such as spectral weight transfer, Fermi surface smearing, to name a few, which we found to have some apparent similarities with different experimental observations on the finite density fermion system. We expect that this search for exotic holographic phases and their experimental realisation will be helpful in understanding the nature of duality between gravitational fields and their associated dual operators.    

Along the above line of discussion, it may be mentioned here that understanding some of the results from condensed matter experiments, such as those of Angle-resolved photoemission spectroscopy (ARPES) \cite{Damascelli:2003bi,2011PhRvL.106l7005K} is a hard task for a theoretical physicist. ARPES experiment reveals the properties of the electronic band structure, quasi-particle excitation, the nature of the Fermi surface (FS) of many topological materials and high $T_c$ superconductors. 
The discovery of superconductivity in the LaBaCuO ceramics at $30~ K$ has opened a new horizon in the field of high-$T_c$ superconductivity \cite{bednorz}.

An unconventional phase of this system which is puzzling physicists over the decades is the pseudo-gap in the fermionic spectral function, which appears as a 

truncated Fermi surface termed as a Fermi Arc \cite{1998Natur.392..157N, 2006PhRvB..74v4510Y, Cremonini:2018xgj, Seo:2018hrc,PhysRevB.99.161113}. 
The appearance of Fermi arc has also been observed in the ARPES experiments in different real condensed matter system such as recently discovered topological Dirac and Weyl semi-metals \cite{PhysRevB.99.161113,YangL,brillaux2020Fermi,Su-Yang,Mehdi,MZHazan,Binghai,PhysRevB.83.205101} and a recent review can be found in \cite{BaiqingLv}. One of the common properties of all the systems mentioned is that the Fermi arc appears in ($2+1$) dimensions either because of the underlying planar structure of the superconductor or as surface states of the bulk material for Dirac and Weyl semi-metals. This phenomenon possesses a mystery, mainly because of the unconventional electronic properties, which can not be explained by usual Landau's theory. Furthermore, the systems are generically shown to be strongly coupled, and this is where the holographic method plays the role.

In the last few years, lots of research were performed to understand the underlying mechanism behind these properties of the Fermi surface from the perspective of condensed matter physics \cite{,PhysRevB.76.174501,PhysRevB.73.174501,PhysRevB.86.115118,PhysRevB.74.125110}, and the same has been done using the holographic method \cite{Hartnoll:2009sz, Herzog:2009xv, McGreevy:2009xe, Zaanen:2015oix, Hartnoll:2016apf}. In some recent pioneering works \cite{Vanacore:2015poa, Vegh:2010fc, Benini:2010qc}, investigations were done to understand the evolution of Fermi arc-like structures, taking RN-AdS black hole and other holographic lattices as their gravity background. 

In this paper, we will study a class of holographic models, where the appearance of the Fermi arc is a generic feature. We will essentially generalize our previous construction \cite{Chakrabarti:2019gow}, where bulk dipole coupling a bulk scalar field that can break the translational symmetry in the boundary. Our goal is to look at the evolution of the Fermi surface and their band structure, with particular emphasis on their observational significances.   

Explicit translational symmetry breaking solutions of a holographically dual \cite{horo1, horo2, horo3,donos4,Donos:2013eha} theory has been studied either by considering spatially periodic chemical potential at the boundary or especially periodic scalar field in the AdS background.  
 
The specific framework we will consider here is Q-lattice  \cite{Donos:2013eha} where holographic lattice background is constructed by exploiting the solution of a complex scalar field in the bulk.
Fermionic spectral function has been studied on the Q-lattice background \cite{Ling2014} together with a dipole type of coupling in the fermion sector. It is to be noted that the secondary Fermi surfaces were noted long back in \cite{SGubser:2020}. In that work, beginning with a dilatonic black hole derived from a truncation of type IIB supergravity, fermionic Green's function dual to massless fermions in the bulk were studied. It was shown that the for not too small bulk fermion charges, there are Fermi surfaces. The Fermi momenta in this case are equally spaced, and there are a finite number of them, proportional to the charge of the bulk fermion. Similar studies using the anisotropic holographic backgrounds have been done in \cite{Iliasov:2019pav,Hyun-Sik:2020,Cremonini2019,Balm2020}. Our goal would be to look into the nature of the holographic Fermi surface evolution due to bulk coupling parameters which will correspond to different dual controlling operators.

We organise the paper as follows: In the next section, we briefly review the Q-lattice background solution, followed by section 3, where we write down the fermion's action and the equations of motion with our definition of the spectral function $\rho\,(\omega,\,\vec{k})$. In section 4, we present the results for different scenarios for massless fermion and discuss each of them. For completeness, in section 5 we discuss the scenarios where by taking non-zero fermion mass. Finally, in the last section, we conclude with the interpretation of our results along with a brief discussion on future directions.

\section{The background geometry }
\begin{figure}[h!]
	\begin{center}
		\includegraphics[width=\columnwidth]{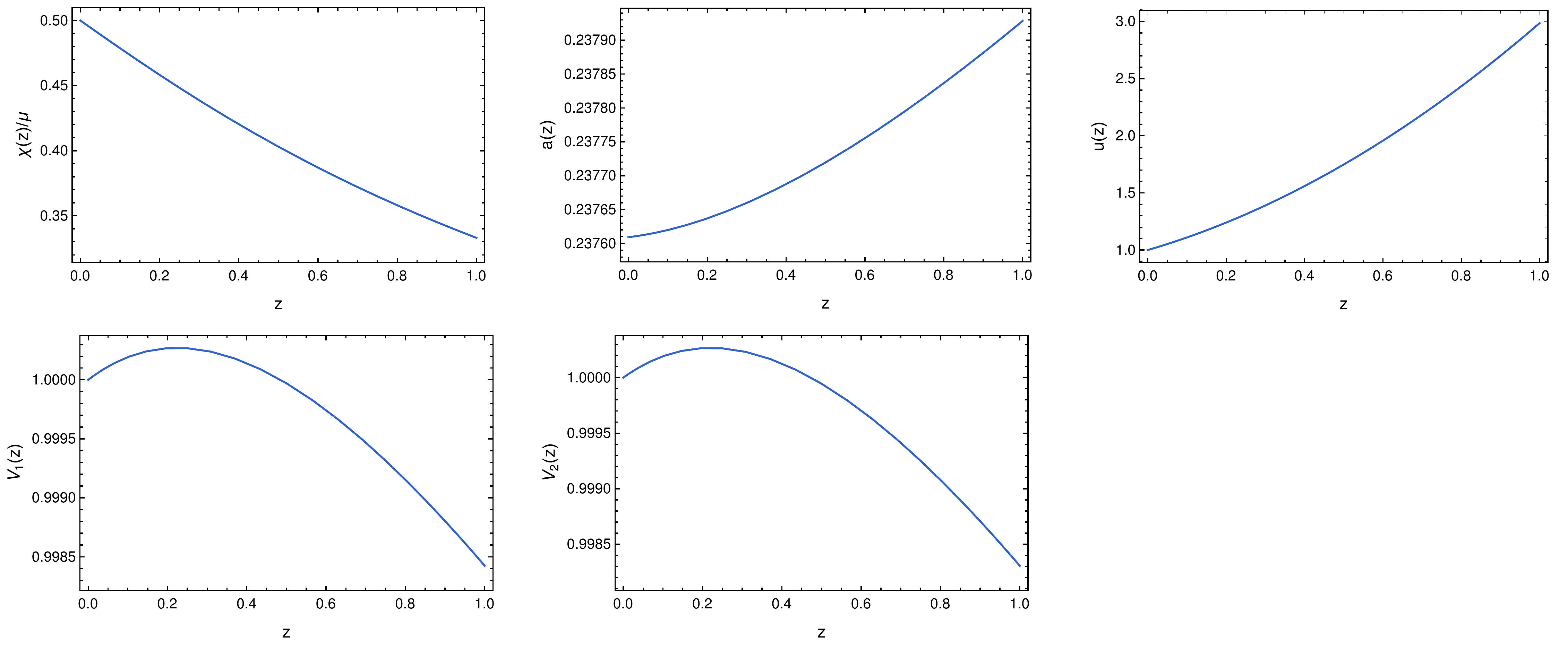}	
		\caption{Q-lattice profile with $m^2_{\phi}=-2$, for the parameters $T/\mu\text{=1}$,\,$\chi^{(1)}/\mu\text{=0.5},k_1/\mu =1\text{ and }k_2/\mu=0$}
		\label{figure1}
	\end{center}
\end{figure}

As has already been mentioned in the introduction, Fermi arc behaviour has been generically observed in two spatial dimension, we consider our holographic system to be in ($2+1$) dimensions. Hence, we shall take a ($3+1$)-dimensional Einstein-Scalar-Maxwell system as a background set up,
\begin{align}\label{action}
	\mathcal{S}=\frac{1}{\kappa^2} \int d^{3+1}x\sqrt{-g}\left[\mathcal{R}+\frac{6}{L^2}-\frac{1}{4}F^2 -|\partial \phi|^2-m_{\phi}^2|\phi|^2\right]
\end{align}
where $\mathcal{R}$ is a Ricci scalar, $L$ being the AdS radius which we will set to be unity later and, Maxwell's field strength $F=dA$, where $d$ is the exterior derivative acting on the vector potential one form $A$. Finally, $\kappa^2=16\pi G$ is the effective reduced gravitational constant which we set to be unity too. As mentioned we consider a complex scalar field $\phi$ which will source the break down of translational invariance of the boundary field theory.
From the above action (\ref{action}) we have the following equations of motions
\begin{align}\label{eqnom}
	&R_{\mu\nu}=\nonumber\\ &g_{\mu\nu}\left(-3+\frac{m_{\phi}^2}{2}|\phi|^2\right)+\partial_{(\mu}\phi\partial_{\nu)}\phi^*
	+\frac{1}{4}\left(2F_{\mu\nu}^2-\frac{1}{2}g_{\mu\nu}F^2\right)\nonumber\\
	&\nabla_{\mu}F^{\mu\nu}=0\;, \;\;\left(\nabla^2-m_{\phi}^2\right)\phi=0 .
\end{align}
We take the ansatz for the metric and the complex scalar field to be of the following form
\begin{align}\label{ansatz}
	&ds^2=-g_{tt}(z)\;dt^2+g_{zz}(z)\;dz^2+g_{xx}(z)\;dx^2+g_{yy}(z)\;dy^2.\\
	&\text{where},\nonumber\\
	&g_{tt}(z)=\frac{U(z)}{z^2};\;g_{zz}(z)=\frac{U(z)^{-1}}{z^2};\;g_{xx}(z)=\frac{V_1(z)}{z^2};\nonumber\\
	&g_{yy}(z)=\frac{V_2(z)}{z^2},\;\;A=(1-z)\;a(z) dt,\nonumber\\
	&\;U(z)\,=\,(1-z)\;u(z),\,\,\phi=e^{i\,k_1\,x+i\,k_2\,y}\;\chi (z)\,.
\end{align}
With the above choice of ansatz for the scalar field, the breaking of translational symmetry by the same scalar field is both in $x$ and $y$ directions.
Here, $u,V_1,V_2,a,\chi$ are all functions of radial coordinate $z$  and $k_1,\,k_2$ are constants interpreted as wave numbers of the lattice. From the above equation we have four second order coupled ODEs for $V_1,\,V_2,\,a,\,\chi$ and one first order for $u$. For a general mass $m_{\phi}^2$, the complex scalar field near the AdS boundary ($z\longrightarrow 0$) behaves as
\begin{align}
	&\chi(z)\,=\,z^{\alpha_-}\,\chi^{(1)}\,+\,z^{\alpha_+}\,\chi^{(2)}\,+ ...
\end{align} 
where, $\alpha_{\pm}=3/2\pm\sqrt{9/4+m_{\phi}^2}$\,. The leading term $\chi^{(1)}$ is associated with the source of the dual scalar operator in the boundary theory, whose dimension is $\Delta=3-\alpha_-=\alpha_+$. To this end we study in detail the case when, scalar field mass $m_{\phi}^2=0$, which corresponds to $\alpha_-=0$ and the corresponding marginal dual scalar operator in the (2+1)- dimensional boundary theory. For building a numerical black hole solution, we solve these equations using the regularity conditions near the horizon $z=1$ and, at the AdS boundary or UV, we assume the following series expansion,
\begin{align}
	&u=1+O(z)\,\,,\,\,V_1=1+O(z)\,\,,\,\,V_2=1+O(z)\;,\nonumber\\ &a= \mu + O(z)
\end{align} 
To solve the non-linear equations first, we linearize the ODEs by discretizing simultaneously both the equations of motions and the boundary conditions following \cite{Trefethen,Boyd} and then use the iterative Newton-Raphson's (NR) method  \cite{Andrade:2017jmt,Krikun:2018ufr} in Mathematica. Moreover for fixed $m^2_{\phi}$, we found that the solutions are specified by 3-dimensionless parameters namely, $T/\mu\,,\, k/\mu$ and  $\chi^{(1)}/\mu^{\alpha_-}$. We notice that for very low temparature (small $T/\mu$), NR method does not converge as it is sensitive to the initial guess values or seeds. We can overcome this by using the solutions at higher temparature as a seeds to achieve another solution at lower temparature.
We illustrate the background solutions in \figurename{ \ref{figure1}}. We consider this as our numerical background in our subsequent analysis for the fermion spectral function. Most of our analysis will be for both fermion and scalar mass to be zero. In this sense our ($3+1$) dimensional gravitational bulk contains  Weyl fermion. Our goal would be to understand the fermion spectral function dual to those Weyl fermion. This could resemble the class of Weyl semi-metals in the real condensed matter systems as pointed out in the introduction.  

\section{Fermions: Action and the spectral function.}

Now that we have the solutions of the background, we can write down the action for fermions with four non-trivial interaction terms given by
\begin{equation}\label{Fermionaction}
	\mathcal{S}_{fer}=\int d^4x\sqrt{-g}i\bar{\psi}\bigg(\slashed{D}-m_{\psi}-i\,\wp_1\slashed{F}-i\,\wp_2\,|\phi|^2\slashed{F}\bigg)\psi\;.
\end{equation}
with $\phi$\; and $m_{\psi}$ being the complex scalar field from the background and fermion mass respectively. We have considered two  coupling parameters: one is simple dipole  coupling parametrised by $\wp_1$ and another is similar to dipole type coupling with the scalar field parametrised by $\wp_2$. At this point, it is to be mentioned that these two types of couplings in the holographic context has already been studied in understanding the Mott-Insulator \cite{Edalati:2010ww} transition or appearance of Fermi Arc \cite{Vanacore:2015poa}. One can interpret the Fermi arc in the present context as the emergence of Mottness  anisotropically due to Lorentz violating couplings. In our previous paper \cite{Chakrabarti:2019gow}, we have further generalised and explored the influence of Lorentz violating dipole coupling on the fermionic spectral function adding a scalar field as a source term. We have constructed the solution associated with the transition from circular to arc like Fermi surface controlled by the scalar condensation in the bulk. In this paper, we look into such system with two sets of couplings and also introduced one dimensional lattice in the boundary theory. For the present case, we will mostly concentrate on the simultaneous effect of those coupling parameters on the fermionic spectral function. Various symbols in (\ref{Fermionaction}) are given by:\;
\begin{align*}
	&\slashed{D}=e^{\mu}_c\Gamma^c\left(\partial_{\mu}+
	\frac{1}{4}\omega_{\mu}^{ab}\Gamma_{ab}-iq A_{\mu}\right)\nonumber\\
	&\slashed{F}=\frac{1}{2}\Gamma^{ab}e^{\mu}_ae^{\nu}_bF_{\mu\nu}\;.
\end{align*}
where,  $e^{\mu}_a ,\omega_{\mu}^{ab}$    are the vielbeins and spin connection, $q$ is the charge of the bulk fermions. 
The Dirac equation from the above action is given by:
\begin{eqnarray}
	\label{DiracEquation1}
	\left(\slashed{D}-m_{\psi}-i\,\wp_1\slashed{F}-i\,\wp_2\, |\phi|^2\slashed{F}\right)\psi=0.
\end{eqnarray}
In order to further simplify the Dirac equation, we choose the following gamma matrices
\begin{align}
	\label{GammaMatrices}
	&\Gamma^{\underline{z}} = \left( \begin{array}{cc}
		-\sigma^3  & 0  \\
		0 & -\sigma^3
	\end{array} \right), \;\;
	\Gamma^{\underline{t}} = \left( \begin{array}{cc}
		i \sigma^1  & 0  \\
		0 & i \sigma^1
	\end{array} \right),\nonumber\\
	&\Gamma^{\underline{x}} = \left( \begin{array}{cc}
		-\sigma^2  & 0  \\
		0 & \sigma^2
	\end{array} \right)\;\;, \;\;
	\Gamma^{\underline{y}} = \left( \begin{array}{cc}
		0  & \sigma^2  \\
		\sigma^2 & 0
	\end{array} \right)
\end{align}

Now, we can expand the Dirac equation as 
\begin{align}
	\bigg(e^{z}_{\underline{z}}\,&\Gamma^{\underline{z}}\,\partial_{z}+S.C+e^{t}_{\underline{t}}\,(\Gamma^{\underline{t}}\,\partial_{t}-iq A_{t})+e^{x}_{\underline{x}}\Gamma^{\underline{x}}\,\partial_{x}+e^{y}_{\underline{y}}\Gamma^{\underline{y}}\,\partial_{y}\nonumber\\&-m_{\psi}-i\,\wp_1\slashed{F}-i\,\wp_2\, |\phi|^2\slashed{F}\bigg)\;\psi=0
\end{align}
where $S.C$ is a contribution from the spin connection that can be cancelled by rescaling the field $\psi=(g_{tt}g_{xx}g_{yy})^{-\frac{1}{4}}e^{-i\omega t + ik_x x + ik_y y}\xi\,(z,\vec{k})$, where $\vec{k}\equiv(-\omega,k_x,k_y)$. Then we have
\begin{align}
	\bigg(&\frac{1}{\sqrt{g_{zz}(z)}}\;\Gamma^{\underline{z}}\;\partial_{z}+\frac{1}{\sqrt{-g_{tt}(z)}}\;\left(\Gamma^{\underline{t}}\;(-i\omega)-iq A_{t}\right)+\nonumber\\&\frac{1}{\sqrt{g_{xx}(z)}}\;\Gamma^{\underline{x}}\;(ik_x)+\frac{1}{\sqrt{g_{yy}(z)}}\;\Gamma^{\underline{y}}\;(ik_y)-m_{\psi}\;+\nonumber\\&\frac{\partial_{z}A_t}{\sqrt{-g_{zz}(z)g_{tt}(z)}}\left(-i\,\wp_1\;\Gamma^{\underline{z}\;\underline{t}}-i\,\wp_2\; |\phi|^2\;\Gamma^{\underline{z}\;\underline{t}}\right)\bigg)\,\xi(z,\textbf{k})=0
\end{align}
Using the basis (\ref{GammaMatrices}) and by writing the spinors $\xi=(\xi_1,\xi_2)^T$, on further splitting as $\xi_j=(\beta_j,\alpha_j)^T$, where $j=1,2$, we have the following radial equations  
\begin{align} 
	&
	\left(\frac{1}{\sqrt{g_{zz}}}\partial_{z}\pm m_{\psi} \right)\left( \begin{matrix} \beta_{1} \cr  \alpha_{1} \end{matrix}\right)
	\mp \frac{(\omega+ q A_{t})}{\sqrt{-g_{tt}(z)}}\left( \begin{matrix} \alpha_{1} \cr  \beta_{1} \end{matrix}\right)
	+\frac{k_x}{\sqrt{g_{xx}}}\left( \begin{matrix} \alpha_{1} \cr  \beta_{1} \end{matrix}\right)\nonumber\\&-\frac{k_y}{\sqrt{g_{xx}}}\left( \begin{matrix} \alpha_{2} \cr  \beta_{2} \end{matrix}\right)
	+\frac{\partial_{z}A_{t}}{\sqrt{-g_{zz}g_{tt}}}\;(\wp_1+\wp_2\;|\phi|^2)\left( \begin{matrix} \alpha_{1} \cr  \beta_{1} \end{matrix}\right)
	=0
\end{align}
\begin{align} 
	&
	\left(\frac{1}{\sqrt{g_{zz}}}\partial_{z}\pm m_{\psi} \right)\left( \begin{matrix} \beta_{2} \cr  \alpha_{2} \end{matrix}\right)
	\mp \frac{(\omega+ q A_{t})}{\sqrt{-g_{tt}(z)}}\left( \begin{matrix} \alpha_{2} \cr  \beta_{2} \end{matrix}\right)
	-\frac{k_x}{\sqrt{g_{xx}}}\left( \begin{matrix} \alpha_{2} \cr  \beta_{2} \end{matrix}\right)\nonumber\\&-\frac{k_y}{\sqrt{g_{xx}}}\left( \begin{matrix} \alpha_{1} \cr  \beta_{1} \end{matrix}\right)
	+
	\frac{\partial_{z}A_{t}}{\sqrt{-g_{zz}g_{tt}}}\;(\wp_1+\wp_2\;|\phi|^2)\left( \begin{matrix} \alpha_{2} \cr  \beta_{2} \end{matrix}\right)
	=0
\end{align}
From these equations, we can expand near the horizon and one can find that the leading terms from the equations of motions are  given by 
\begin{eqnarray}
	\label{horizon}
	&&
	\partial_{z}\left( \begin{matrix} \beta_{j}(z,\vec k) \cr  \alpha_{j}(z,\vec{k}) \end{matrix}\right)
	\mp \frac{\omega}{4\pi T}\frac{1}{1-z}
	\left( \begin{matrix} \alpha_{j}(z,\vec{k}) \cr  \beta_{j}(z,\vec{k}) \end{matrix}\right)
	=0.
\end{eqnarray}
From this equation, we determine the in-falling boundary condition at the horizon for extracting the retarded Green's function at the AdS  boundary, and the independent ingoing boundary condition should be imposed at the horizon, i.e.,
\begin{equation}
	\left( \begin{matrix} \beta_{j}(z,\vec{k}) \cr  \alpha_{j}(z,\vec{k}) \end{matrix}\right)
	\sim c_j(1-z)^{-\frac{i\omega}{4\pi T}}.
\end{equation}
Furthermore, the asymptotic behaviour of the Dirac equations near the AdS boundary $(z\rightarrow 0)$ has the following form
\begin{equation} \label{}
	\left( \begin{matrix} \beta_{j} \cr  \alpha_{j}\end{matrix}\right)
	{\approx}\; A_{j}z^{m_{\psi}}\left( \begin{matrix} 1 \cr  0 \end{matrix}\right)
	+D_{j}z^{-m_{\psi}}\left( \begin{matrix} 0 \cr 1 \end{matrix}\right).
\end{equation}
We can numerically read off the coefficients $A$ and $D$ to obtain the Green's functions.
For the couplings considered here, we have a mixing of various spinorial components. We followed the usual prescription  used in \cite{Faulkner:2009am,Guarrera:2011my, Vanacore:2015poa} to extract the Green's
function by using two sets of linearly independent boundary conditions that are given by
\begin{align}\label{defn}
	\left(\begin{array}{cc}\beta^I_1& \beta^{II}_1\\ \beta^I_2 & \beta^{II}_2 \end{array} \right)=  \left(\begin{array}{cc} s_{11}& s_{12}\\ s_{21} & s_{22} \end{array} \right) \left(\begin{array}{cc}\alpha^I_1& \alpha^{II}_1\\ \alpha^I_2 & \alpha^{II}_2 \end{array} \right),
\end{align}
Now the retarded Green's function is defined as   
\begin{align}\label{GreenFunction}
	G^R(\omega,k_x,k_y)=-i \left(\begin{array}{cc} s_{11}& s_{12}\\ s_{21} & s_{22} \end{array} \right)\cdot \gamma^t
\end{align}
with gamma matrices in our basis as $\gamma^t=i\sigma_1$.
The observable quantity of interest, is the spectral function $\rho\,(\omega,k_x,k_y) $ which is given by
\begin{align}
	\rho\,(\omega\,,\vec{k})=\text{Im}\left[\text{Tr}\, G^R(\omega\,,\vec{k})\right].
\end{align}
where $\vec{k}\,\equiv\,(k_x,k_y)$ and $ G^R(\omega\,,\vec{k})$ is the retarded Green's function. The properties of $\rho\,(\omega\,,\vec{k}) $ is what we will study in our subsequent analysis.

\section{Results and Discussion}

As already emphasized thorough out the earlier sections, our goal would be to explore the behaviour of the Fermi surface considering various coupling $\wp_1\,,\wp_2$  in a systematic manner.

We wish to discuss the properties of the Fermi surface close to zero temperature. Before we explore different possibilities, as a consistency check, we reproduce the result in  \figurename{ \ref{figure2}} obtained in \cite{Ling2014,Iliasov:2019pav} assuming all the coupling parameters set to zero.
\par 
\begin{figure}[htbp]
	\begin{center}
		\includegraphics[width=\columnwidth]{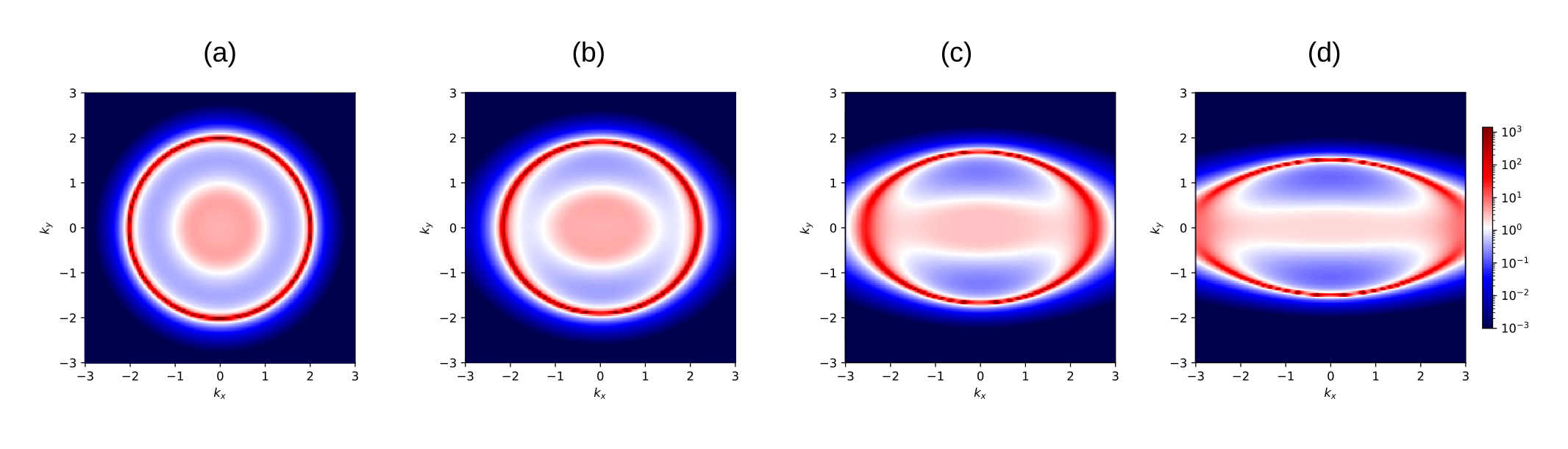}
		\caption{Spectral function $\rho\,(k_x,k_y)$ with all $\wp$'s set to zero. Panel $\bf{(a)-(d)}$ is for $\frac{\chi^{(1)}}{\mu^{\alpha_-}}=0.2,\,1.0,\,2.0,\,3.0$  respectively.  Here $m_{\psi}=0\;,q=1$, background parameters $T/\mu=0.009$,\, $k_1/\mu=0.8.$ and $k_2/\mu=0.$}
		\label{varysourcefixAllP=0}
	\end{center}
\end{figure}
\begin{figure}[htbp]
	\centering
	\includegraphics[width=\columnwidth]{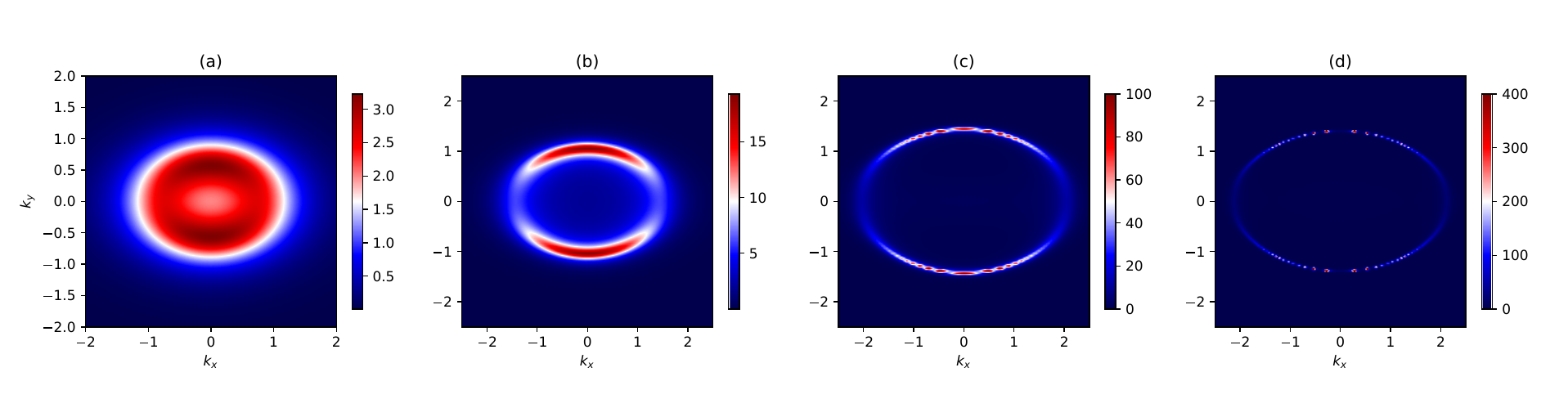}
	
	\caption{Density plot of the spectral function $\rho\,(k_x,k_y)$ with all couplings set to zero for the parameters $m_{\psi}=0$ and $q=1$. In panel (a) to (d) the background parameter $\chi^{(1)}=2.0,\,k_1/\mu=0.8,\,k_2/\mu=0,\,m_{\phi}^2=0$ and $T/\mu=0.09,0.04,0.02,\text{and}\,0.009$ respectively. Clearly as we lowered the temperature the Fermi surface becomes very sharp. }
	\label{figure2}
\end{figure}

\begin{figure}[t]
	\centering
	\includegraphics[width=\columnwidth]{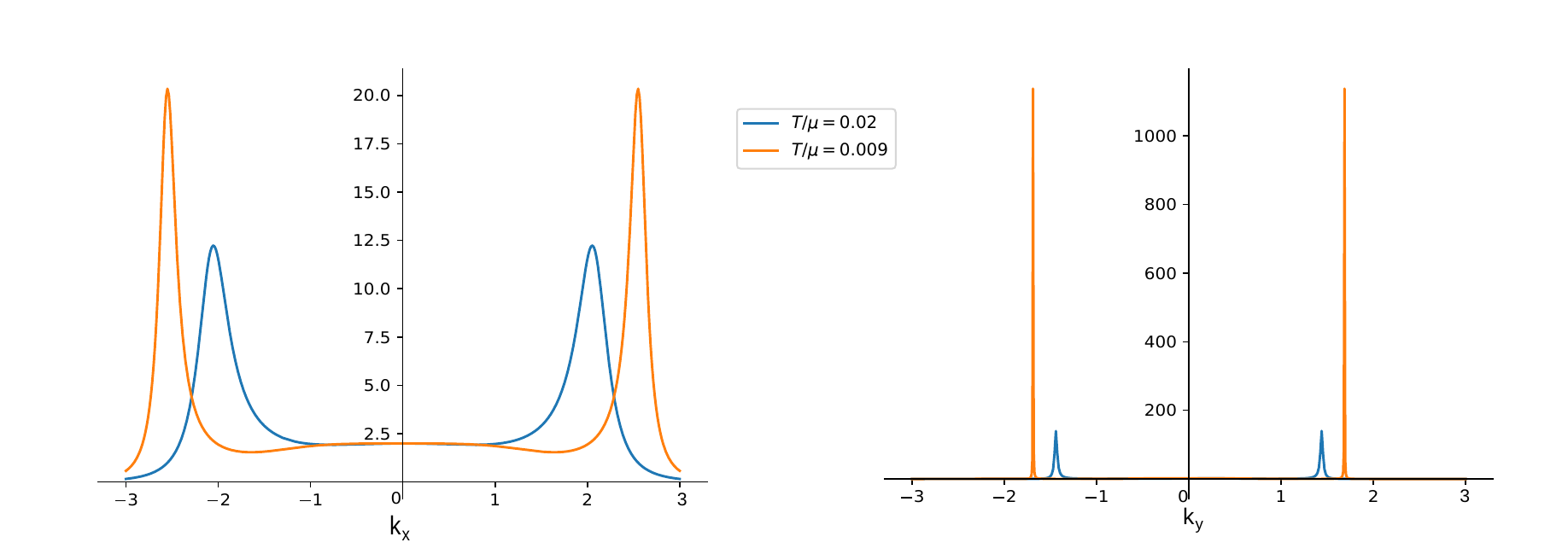}
	
	\caption{2D slice along $k_y=0$ (left) and  $k_x=0$ (right) of \figurename{ \ref{figure2}} to compare the peaks for $T/\mu=0.02$ (blue) and $0.009$ (orange) respectively with other parameter values same as in \figurename{ \ref{figure2}}. }
	\label{figure3}
\end{figure}
\begin{figure}[t]
	\centering
	\includegraphics[width=\columnwidth]{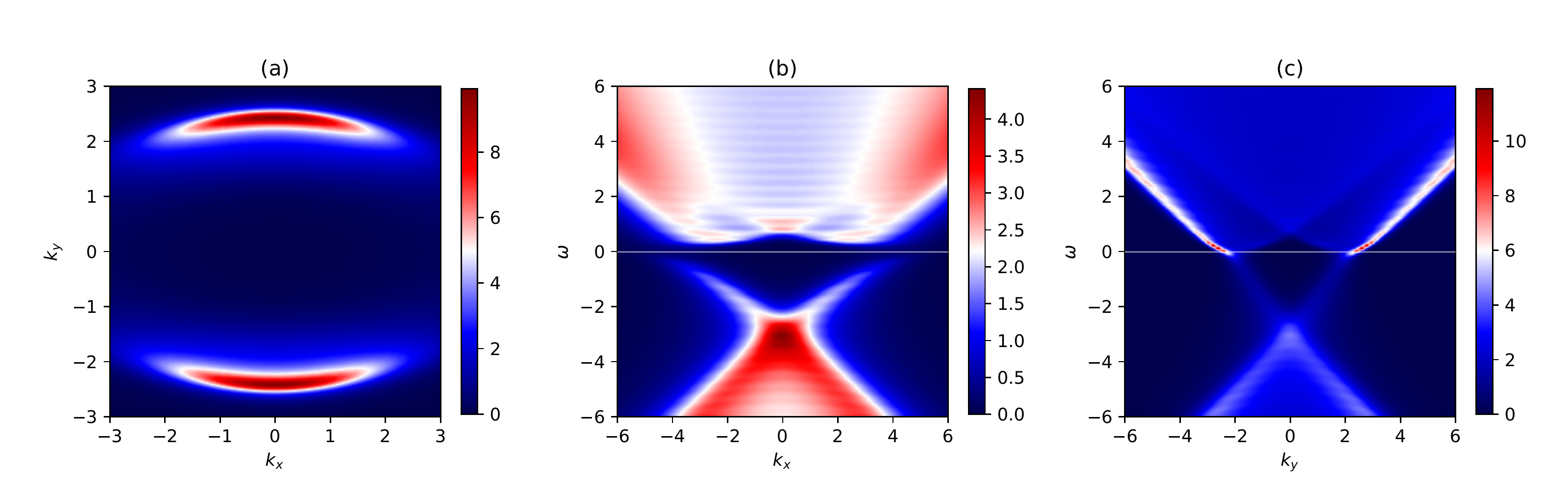}
	
	\caption{Panel $\bf{(a)}$: The fermionic spectral density in the $k_x-k_y$ space for $\wp_1=1$ near $\omega=0$. Coupling $\wp_2$ set to zero. Panel $\bf{(b) \& (c)}$: The energy-momentum distribution along $k_x$ and $k_y$ respectively. Here, the fermion mass $m_{\psi}=0$ and charge $q=1$ with background parameters $\chi^{(1)}=2.0,\,k_1/\mu=0.8,\;k_2/\mu=0,\,m_{\phi}^2=0$ and $T/\mu=0.02\,.$ The thin white line is at $\omega=0$ which is the Fermi level.}
	\label{densityp1}
\end{figure}

\subsection{Symmetry broken along $k_1$ only}
\noindent Firstly, let us set our background parameter $k_2=0$ and consider only $k_1$.
The influence of the background scalar on the spectral function can be seen from  \figurename{ \ref{varysourcefixAllP=0}}, $\chi^{(1)}$ represents the source of the dual boundary operator of the bulk scalar field. As we increase the strength of the boundary source term, the amplitude of the fermionic spectral function reduces along $k_x$. Also in the limit of $\chi^{(1)}\to0$, the spectrum reduces to the one observed in case of RN-AdS blackhole with a circular Fermi surface and no gap (see \figurename{ \ref{varysourcefixAllP=0}}$\; {(a)}$). This suppression can be attributed to reducing a quasi-particle lifetime near the $k_x^F \simeq \pm 2 $ point.
From \figurename{ \ref{figure2}} ($(b)-(d)$), at low temperature one notices that, without any coupling terms, the spectral function gives rise to a stretched Fermi surface with its amplitude suppression along $k_x-$direction $(k_y =0)$, arising due to the broken translational symmetry along $x-$direction established by the background scalar field. Whereas, along $k_y$ direction, the spectral function indicates the existence of the Fermi surface with a large density of states. Further, as one lowers the temperature, the peak of the Greens function increases, which essentially enhances the quasi-particle lifetime.   
On the other hand, turning on the pure dipole coupling  $\wp_1=1$, the Fermi surface is completely destructed along $k_y=0$ line giving rise to dipolar Fermi arcs  \figurename{ \ref{densityp1}}(a). This destruction can be characterized by the opening of the energy gap that can be clearly observed from panel $(b)$ of \figurename{ \ref{densityp1}}. With this consistency check of present calculations with the previous works, we shall now examine the evolution of the spectral function or, in other words, the nature of the Fermi surface due to change of different controlling parameters such as coupling $(\wp_1,\wp_2)$, temperature $(T/\mu)$, and source strength $\chi^{(1)}$ of the scalar field at the boundary. 

\begin{figure}[t]
	\centering
	\includegraphics[width=\columnwidth]{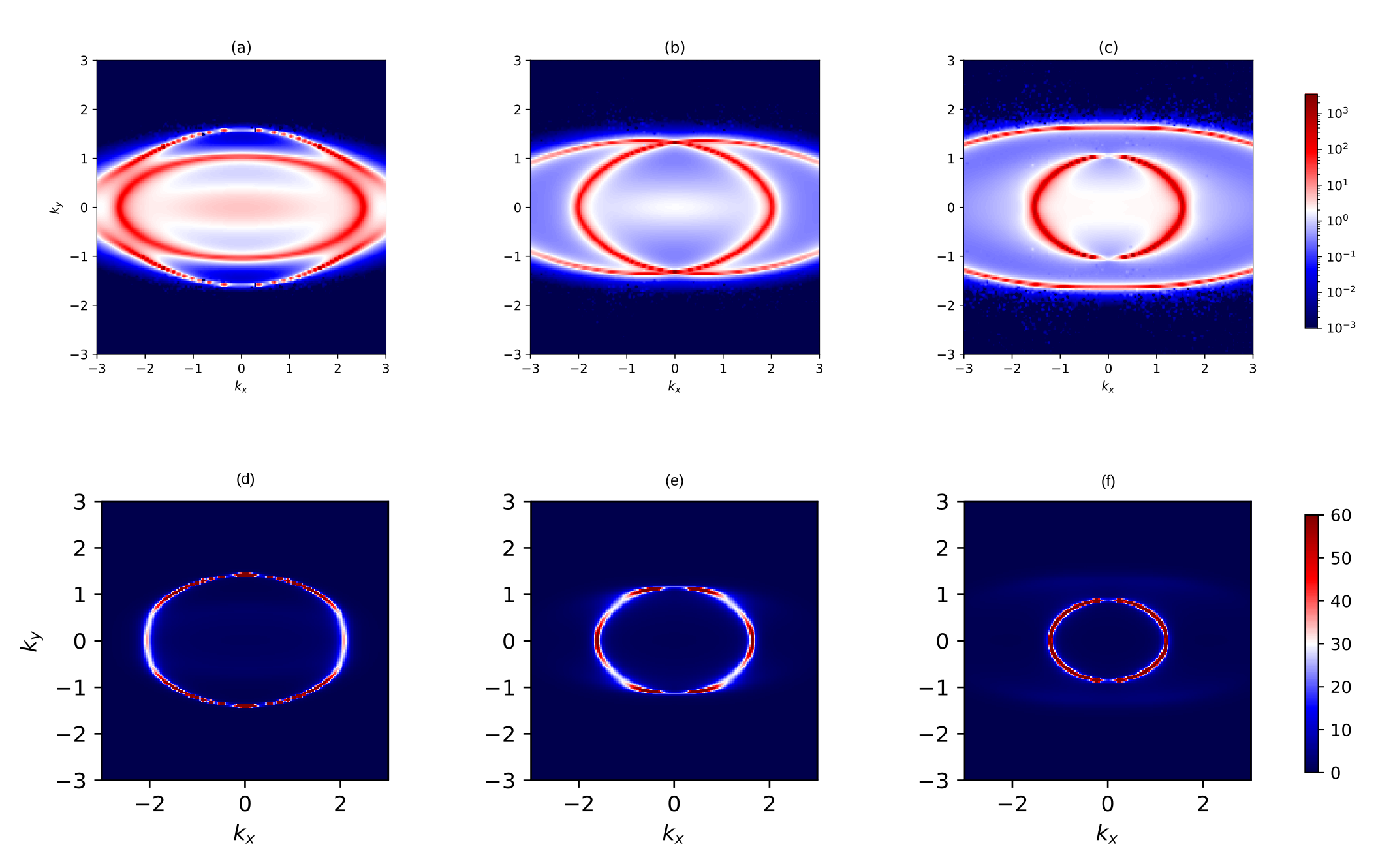}
	\caption{Spectral density  $\rho\,(k_x,k_y)$ for $T/\mu=0.001\,$(top) and $T/\mu=0.009\,$(bottom) with $\chi^{(1)}/\mu^{\alpha_-}=2.0$ and $k_1/\mu=0.8$. $\wp_1$ is set to zero and $\wp_2$ is non-zero and  $m_{\psi}=0\;,q=1$. Both the top  and bottom panel from left to right $\wp_2\,=\,[-0.1,-0.2,-0.3]$.}
	\label{figp1p3}
\end{figure}

\begin{figure}[htbp]
	\centering
	\includegraphics[width=\columnwidth]{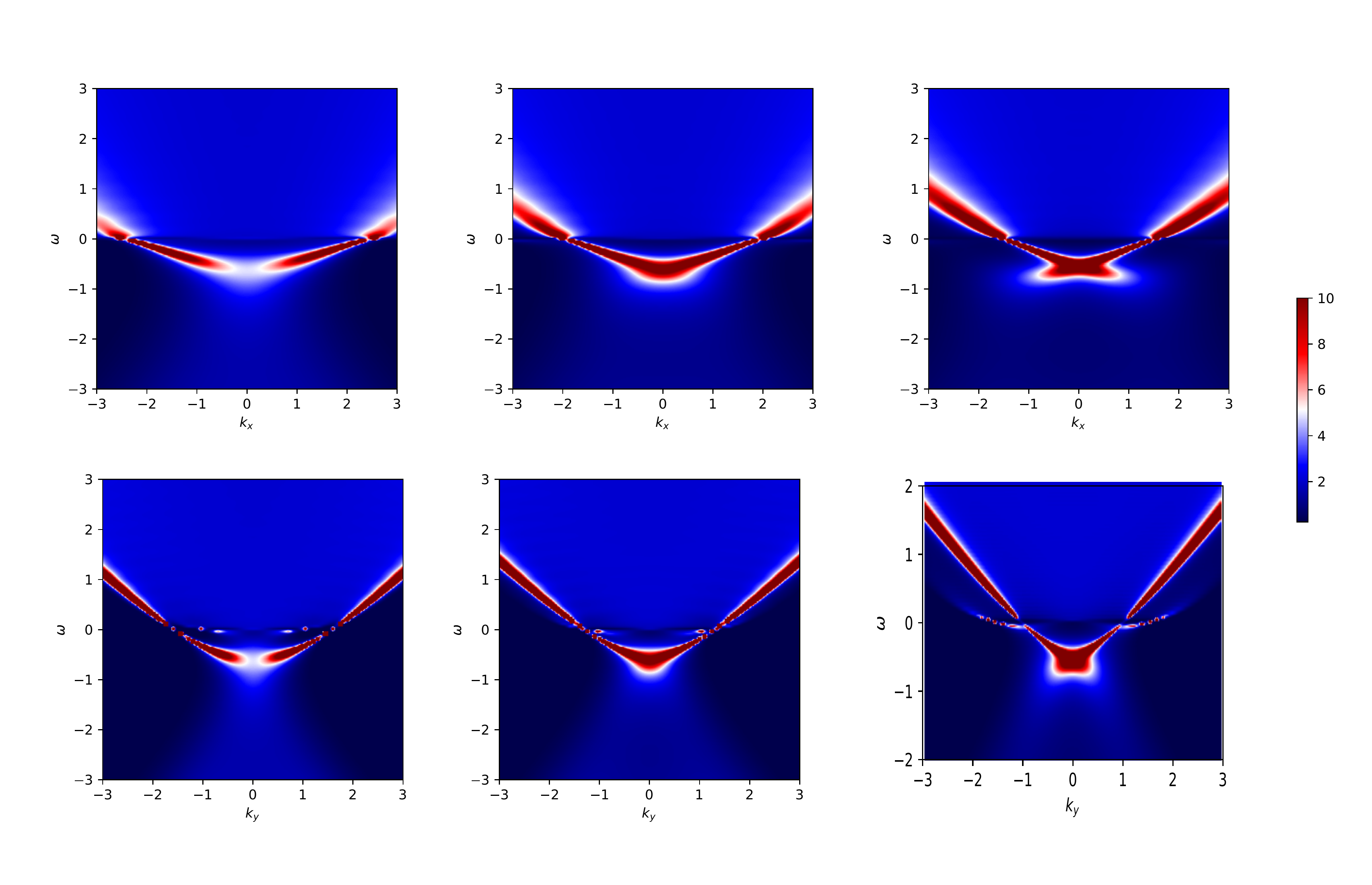}
	\caption{Spectral density  $\rho\,(\omega,\vec{k})$ for $T/\mu=0.001$ with $\chi^{(1)}/\mu^{\alpha_-}=2.0$,\,  $k_1/\mu=0.8$ and $k_2/\mu=0$. Here, parameter $\wp_1=0$ and $\wp_2$ is non-zero and  $m_{\psi}=0\;,q=1$. In the top  and bottom panel from left to right $\wp_2\,=\,[-0.1,-0.2,-0.3]$.}
	\label{omkxkyp3}
\end{figure}

{\bf Controlling both the dipole couplings $(\wp_1, \wp_2)$:} No special features appeared in the spectral function upon increasing the value of $\wp_1$, with $\wp_2=0$, except the known two Fermi arcs becoming more flattened (see  \cite{Ling2014}) and prominent. Similar behaviour is observed for the non-zero but positive value of $\wp_2$, which we chose not to discuss any further here. However, interesting behaviour of the spectral function emerges if the dipole coupling $\wp_2$ assumes a negative value, which is further controlled by the anisotropic scalar field profile in the background. The background anisotropic scalar field plays a crucial role in controlling the shape and the amplitude of the spectral function. As discussed, we will consider a complex scalar field in the bulk that will have a non-normalizable source at the boundary, which can be used as a tunable parameter. For this, we choose the mass of the scalar field to be $m_{\phi} =0$, later on, we will comment on non-zero masses which are outside the $\mbox{AdS}_2$ BF bound.  Nevertheless, as we independently increase the magnitude of the dipole coupling parameter $\wp_2$, a pair of Fermi surfaces emerges as shown in \figurename{ \ref{figp1p3}}. The appearance of multiple Fermi surfaces in holographic model was initially reported in \cite{SGubser:2020} for spherically symmetric background. Also recently in \cite{Cremonini2019,Balm2020}, more complicated background geometry were introduced while breaking  translational invariance and Fermi  arcs like features were observed triggered by the effective mass term and the lattice effects.
The emergence of this pair of  Fermi surfaces that are anisotropic is the interesting feature of our results as they have been observed in the ARPES experiments in different real condensed matter systems such as Topological insulators (TI), recently discovered Dirac and Weyl semi-metals \cite{PhysRevB.99.161113,YangL}. For superconductivity, it is the pseudogap region in the phase diagram where the Fermi arc appears with a controlled doping concentration in a 2- dimensional layer of the superconducting sample, and it is connected to the suppression in the Spectral function. For the case of Dirac and Weyl semi-metals, Fermi arcs appear as surface states which converge at the two distinct Dirac points in the three-dimensional bulk. We also showed the presence of these gappedless surface states in the energy-band dispersion in \figurename{ \ref{omkxkyp3}} with only coupling $\wp_2$ and in \figurename{ \ref{figure8omk}} with the two couplings parameters $\wp_1$ and $\wp_2$. Further, from \figurename{ \ref{figure8omk}}, one can notice the existence of flat band near the Fermi surface. Interestingly such a highly flat band has been observed in the real condensed matter system, such as in the bilayer graphene \cite{DMarchenko}.  The double Fermi arcs for our holographic system can be thought of as holographic surface states of a bulk material located in ($2+1$) dimensional boundary of the AdS space \cite{Eoh}.\par

Further observation in \figurename{ \ref{figp1p3}} $(e)$ at $T/\mu=0.009$,though in the cuprates, the presence of a secondary Fermi surfaces are not yet known  but here it shows that when these two surfaces crossed each other to form a $d-$wave like structure Fermi surface. These phenomena have an interesting resemblance with the usual high-temperature superconductor as discussed in a recent solvable toy model using conventional method \cite{PhysRevB.103.024529}. However, the shrinking of the Fermi surface in the radial momentum direction can be tied with the reduction of the available density of states for the system. This shrinking of the Fermi surface does happen in the pseudogap region for the high-temperature superconducting system as one decreases the temperature in the appropriate range of doping concentration.
 
\begin{figure}[htbp]
	\centering
	\includegraphics[width=\columnwidth]{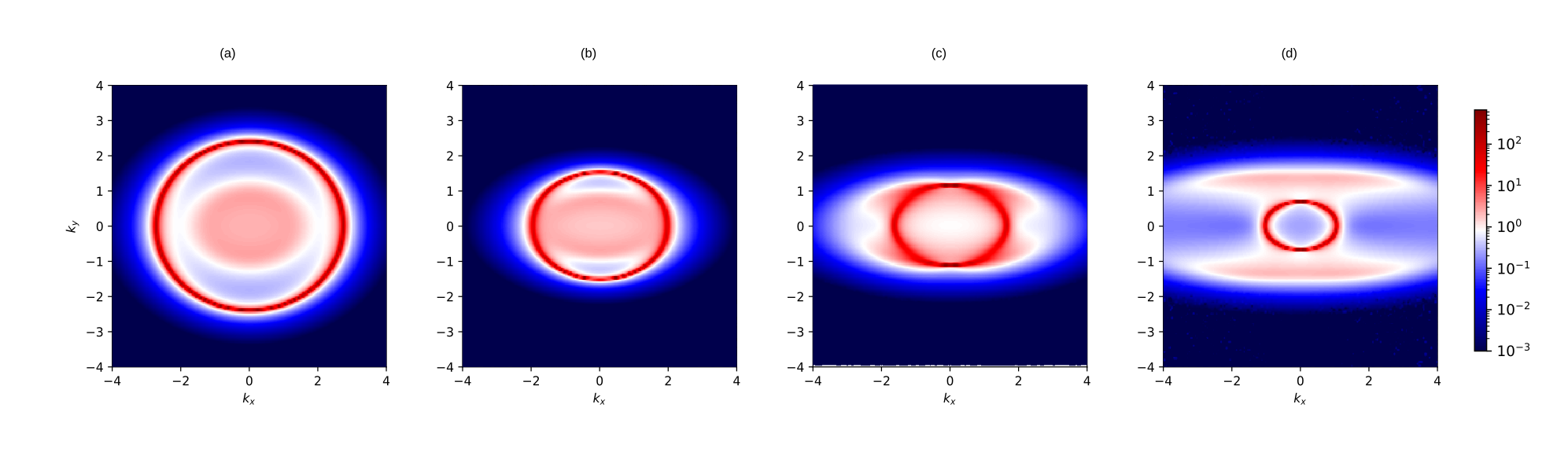}
	\caption{Plot of $\rho\,(k_x,k_y)$ with $\wp_2=-0.2$ and $\wp_1=0$. Panel $(a)-(d)$ is for $\frac{\chi^{(1)}}{\mu^{\alpha_-}}=1.0,\,1.5,\,2.0,\,2.5$  respectively. Here, $m_{\psi}=0\;,q=1$,\,$k_1/\mu=0.8$,\;$k_2/\mu=0$, and\, $T/\mu=0.009$.}
	\label{varysourceP3}
\end{figure}

The same behaviour of the Fermi surface can be observed by keeping fixed $\wp_2=-0.2$, and controlling source at the boundary of the scalar field (see \figurename{ \ref{varysourceP3}}). Therefore, the source identified by the non-normalizable part of the bulk scalar field $\chi^{(1)}$ can be thought of as a doping parameter in a certain boundary field theory. In the \figurename{ \ref{omkxkyp3}}, we plotted the evolution of energy band diagram in both $k_x$ and $k_y$ direction. It clearly captures the opening and closing of the energy gap depending on $\wp_2$ values specifically along $k_y$ direction.   

\begin{figure}[htbp]
	\centering
	\includegraphics[width=\columnwidth]{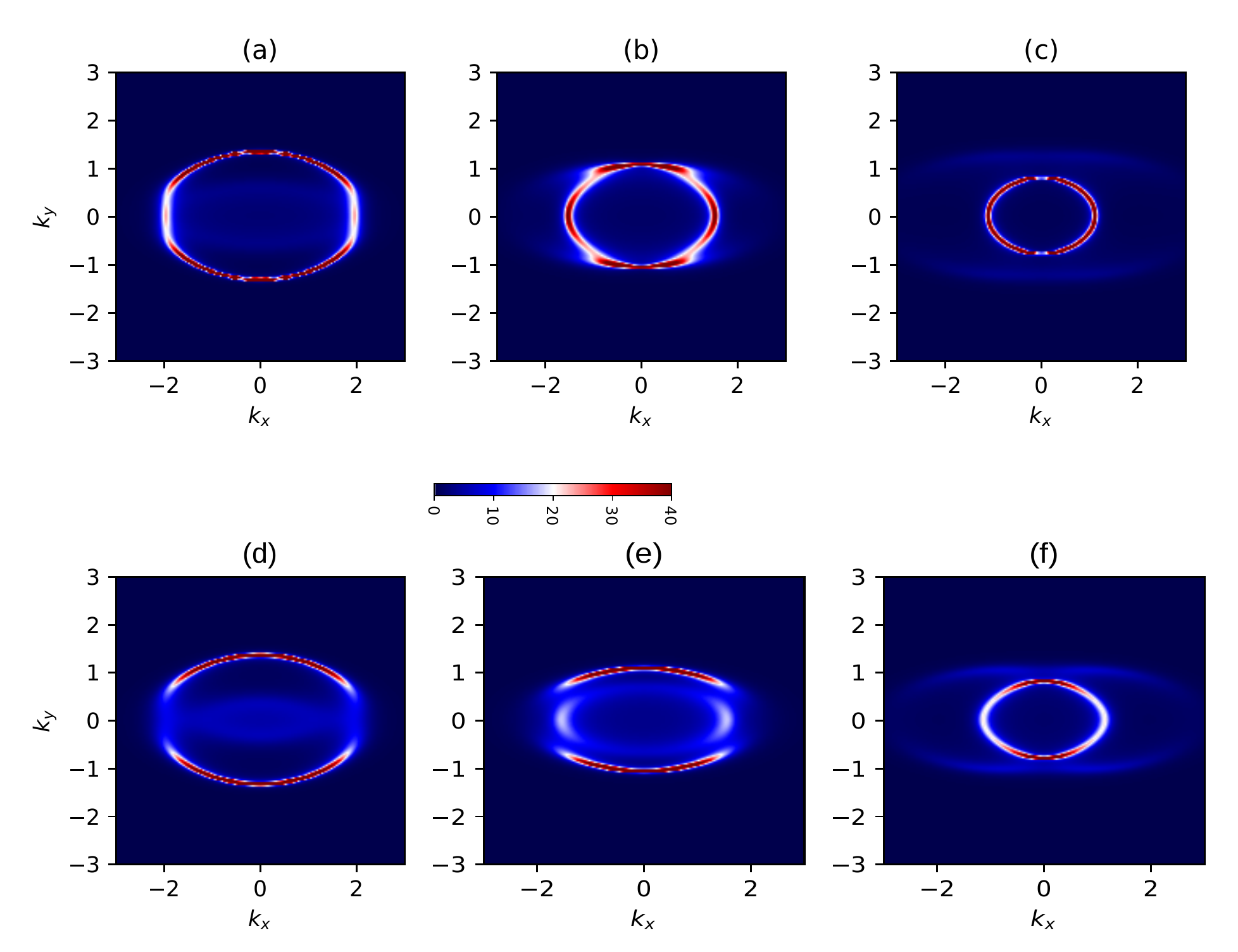}
	\caption{Spectral function $\rho\,(k_x,k_y)$ with $\wp_1$,\, $\wp_2$ set to non-zero and  $m_{\psi}=0\;,q=1$,\, $T/\mu=0.009$\,,\;$\frac{\chi^{(1)}}{\mu^{\alpha_-}}=\,2.0$\,,\;$m_{\phi}^2=0$\,,\;$k_1/\mu=0.8$ and $k_2/\mu=0.0$. Panel (a)-(c) are for  [\,$(\wp_1\,,\wp_2)\,]\,=\,[\,(0.2,-0.2),(0.2,-0.3),(0.2,-0.4)]$ and panel $(d)-(f)$ are for  [\,$(\wp_1\,,\wp_2)\,]\,=\,[\,(0.5,-0.3),(0.5,-0.4),(0.5,-0.5)]$ respectively.}
	\label{figurep1p3}
\end{figure}
\begin{figure}[htbp]
	\centering
	\includegraphics[width=\columnwidth]{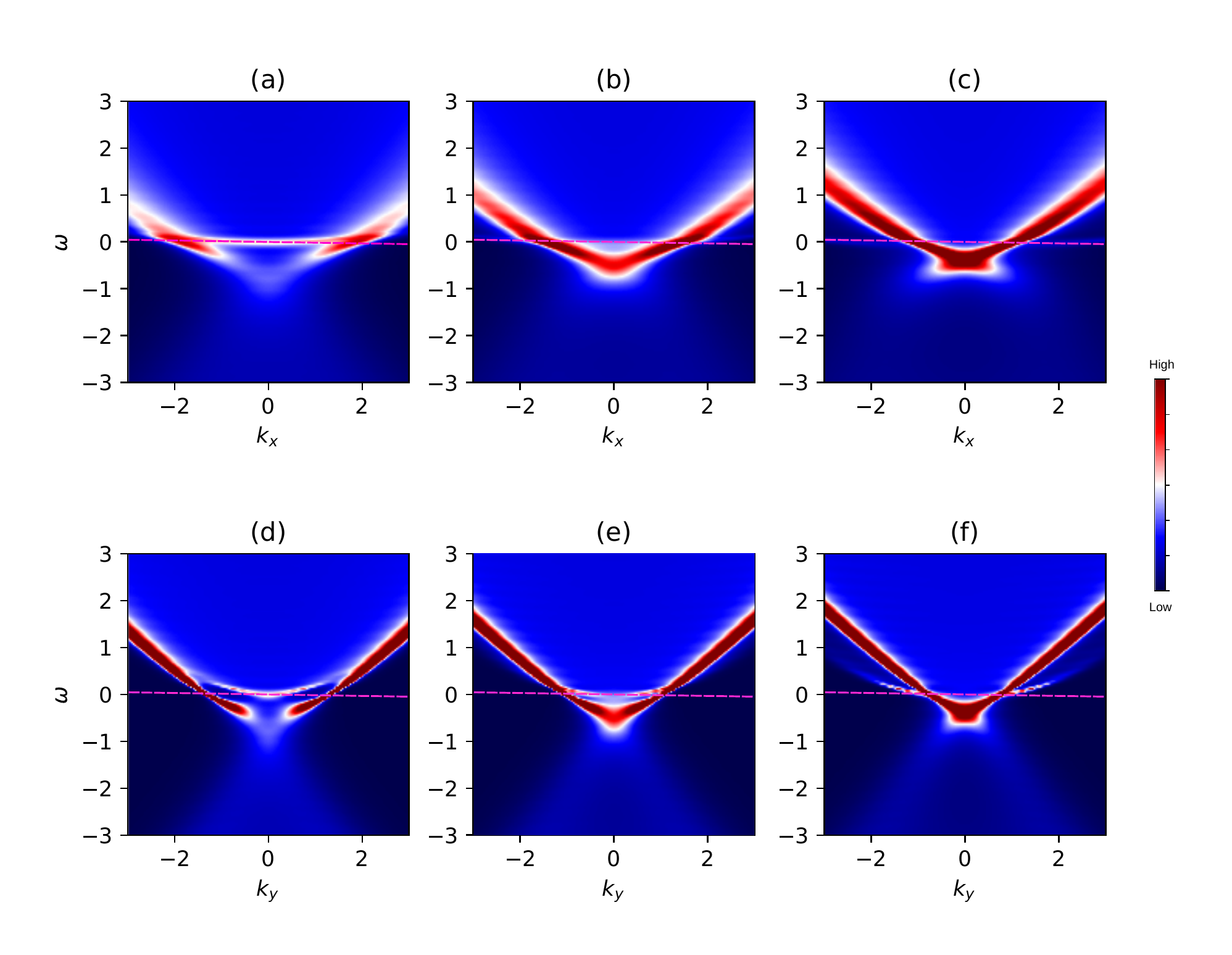}
	\caption{Energy-momentum dispersion  with $\wp_1$,\, $\wp_2$ set to non-zero and  $m_{\psi}=0,\;q=1$,\, $T/\mu=0.009$,\;$m_{\phi}^2=0$, $k_1/\mu=0.8$  and $k_2/\mu=0$. Top and bottom panel are along $\omega-k_x$ and  $\omega-k_y$ respectively. From left to right the parameters [\,$(\wp_1\,,\wp_2)\,]\,=\,[\,(0.5,-0.2),(0.5,-0.3),(0.5,-0.4)]$ .}
	\label{figure8omk}
\end{figure}

So far, we have discussed about controlling the dipole coupling independently. Interesting behaviour of the Fermi surface also emerges (see \figurename{ \ref{figurep1p3}}), once we turn on both the coupling parameters. We assume two sample values of $\wp_1 = 0.2,\,0.5$ , and vary $\wp_2$ across the range $(-0.2, -0.4)$ keeping the temperature fixed at $T/\mu=0.009$, and background translational symmetry breaking parameter $k_1/\mu=0.8$ same. What is observed is that for a given $\wp_1$, at a lower value of $\wp_2$, the outer dipolar Fermi arcs is sharp with an additional closed Fermi surface inside the arc. Interestingly with the increasing $\wp_2$, the outer Fermi arc is gradually disappearing along with the prominent appearance of the inner closed Fermi surface through the transfer of spectral weight from the outer to the inner one.  Such transfer of spectral weight has been observed in the real condensed matter system \cite{transfer} such as La$_{1-x}$Sr$_x$MnO$_3$. ARPES results on those materials shows the changes in the electronic structure across the metal-insulator transition, which is attributed to the gradual disappearance of the Fermi surface near the Mott-Insulator transition. The reason has been given due to transfer of
spectral weight from the coherent band near the Fermi level to the lower Hubbard band. 
Such transfer of spectral weight can also be seen in our holographic energy band diagram shown in \figurename{ \ref{figure8omk}}. Furthermore, at lower $\wp_2$, $\omega =0$ Fermi surface seems to be smeared over a large momentum range. This observation has striking similarities with the phenomena of Fermi surface smearing observed experimentally in disordered condensed matter system \cite{FSsmearing}. The role of bulk dipole parameters can be thought of as the strength of the disorderness in the boundary. As we increase $\wp_2$, the available smeared FS states specifically near $k=0$ move toward the higher momentum state giving rise to specific Fermi momentum. The emergence of such multiple Fermi surfaces can also be understood from a different perspective that the boundary strongly coupled system contains different species of emergent fermionic degrees of freedom corresponding to their intrinsic properties, which must be coupled with the dual operators associated with bulk dipole couplings.  
\subsection{Symmetry broken in both $k_1$ and $k_2$ spatial direction}
\begin{figure}[htbp]
	\centering
	\includegraphics[width=\columnwidth]{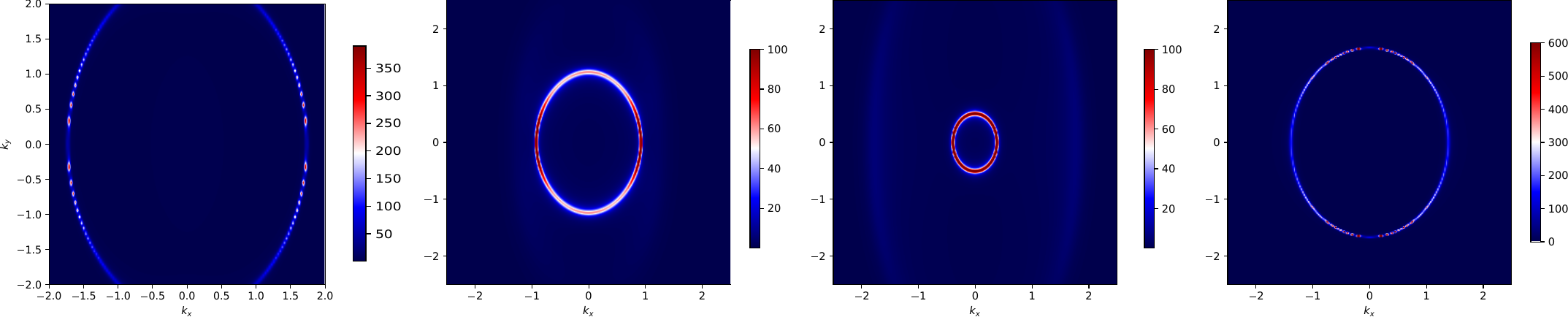}
	\caption{Spectral density with background parameters  $T/\mu=0.009$, $k_1/\mu=0.2$,\;$k_2/\mu=0.8$, $m_{\phi}^2=0$ and, $\chi^{(1)}=2$. Here, the fermion mass $m_{\psi}=0\;,\text{and charge}\;q=1$. Left to right we vary $\wp_2=0,-0.3,-0.5,\;\text{and}\;-1.2$ while fixing $\wp_1=0$. Right most panel showing a zoom in version of the inner Fermi surface.}
	\label{P2k1=0.2k2=0.8varyP2}
\end{figure}

\begin{figure}[htbp]
	\centering
	\includegraphics[width=\columnwidth]{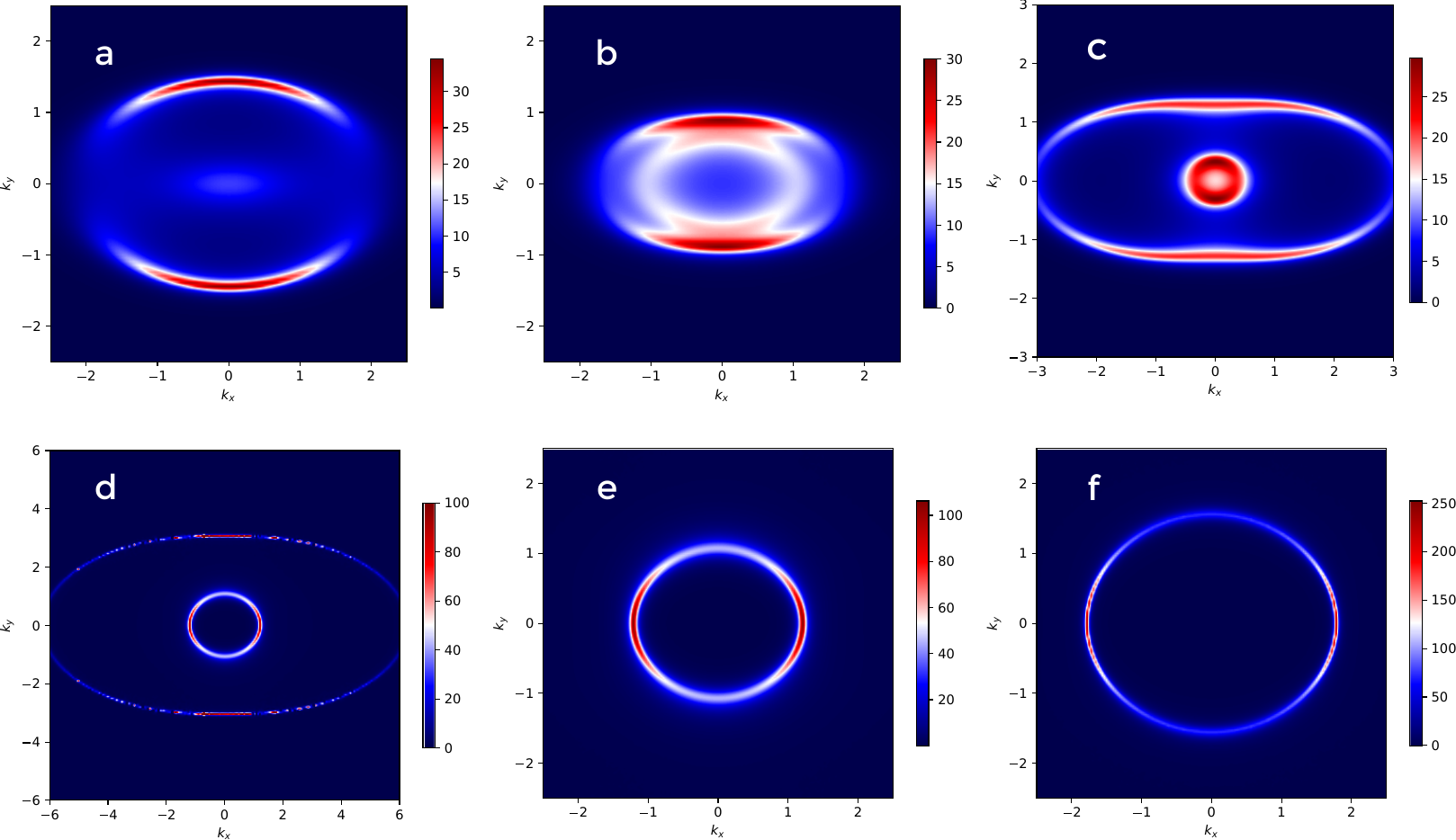}
	\caption{Spectral density with background parameters  $T/\mu=0.009$, $k_1/\mu=0.8$,\;$k_2/\mu=0.2$, $m_{\phi}^2=0$ and, $\chi^{(1)}=2$. Here, the fermion mass $m_{\psi}=0,\;\text{ and charge}\;q=1$. In the panel $(a)-(d)$ we vary $\wp_2=-0.3,-0.5,-0.7,-1.2$  and fixing $\wp_1=0.5$. Panel $(e)$: Zoom of the inner FS in $(d)$ and $(f)$ is a zoom version for $(\wp_1,\wp_2)=(0.5,-1.2)\;\text{and}\,(0.5,-1.4)$ respectively.}
	\label{P1P2k1=0.8k2=0.2varyP2}
\end{figure}

\begin{figure}[htbp]
 	\centering
 	\includegraphics[width=\columnwidth]{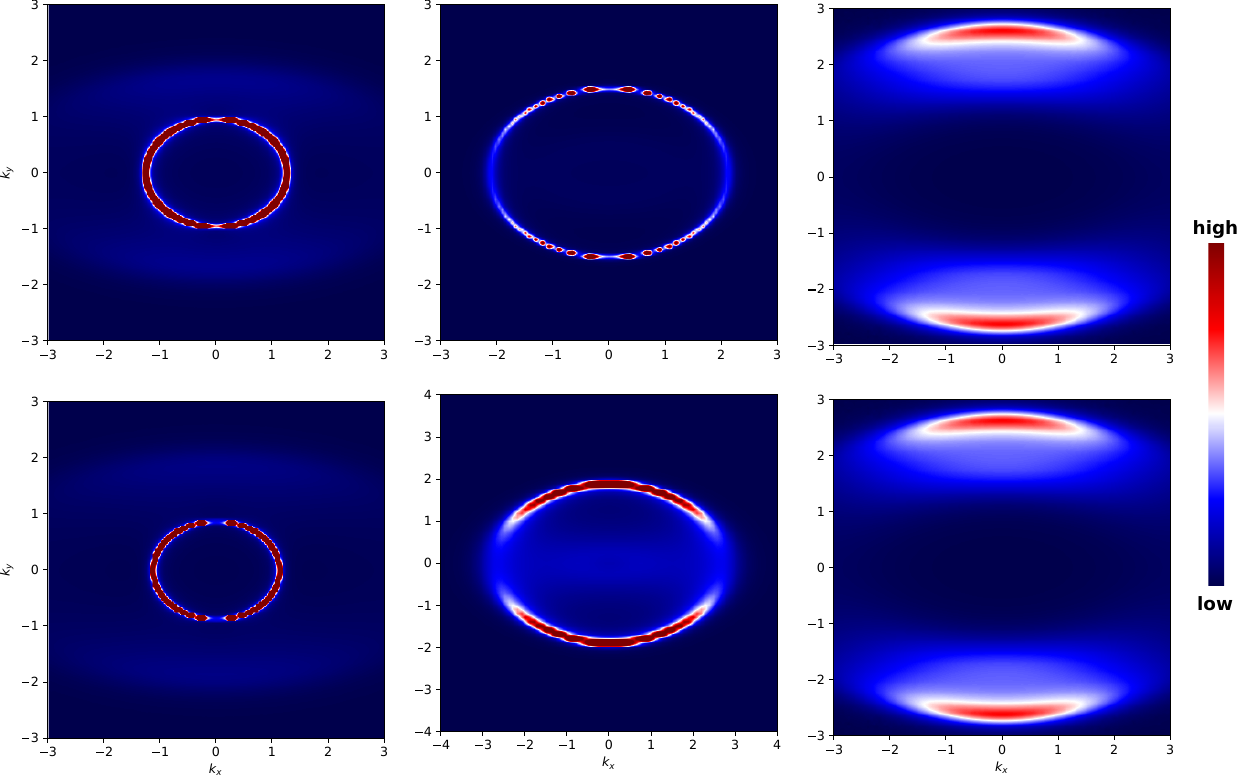}
 	\caption{Spectral density with background parameters  $T/\mu=0.009$, $k_1/\mu=0.8$,\;$k_2/\mu=0.2$, $m_{\phi}^2=0$ and, $\chi^{(1)}=2$. Here, the fermion mass $m_{\psi}=0,\;\text{ and charge}\;q=1$. Top panel is for  combination of $\wp_1$ and $\wp_2$ in the IR while bottom panel is for the UV. From left to right, the effective coupling is negative, zero and positive. }
 	\label{interplayP2}
 \end{figure}

So far we have look into the scenario where translational symmetry is broken only in the $x-$direction. Now let us check what happens when symmetry is broken in both $x$ and $y$ spatial directions. As expected, when $k_2/k_1>1$, the gap along $k_y$  now opens up  as shown in the first panel of \figurename{ \ref{P2k1=0.2k2=0.8varyP2}}. Similar interesting features were observed as shown in  \figurename{ \ref{P2k1=0.2k2=0.8varyP2}} and \figurename{ \ref{P1P2k1=0.8k2=0.2varyP2}}. On zooming in on the inner Fermi surface \figurename{ \ref{P2k1=0.2k2=0.8varyP2}}(e-f), the spectral weight is suppressed in the opposite direction in comparison to that of the outer surface for smaller $\wp_2$ values. However, for $k_1=k_2$ the Fermi surface is circular in shape as in the case of RN-AdS blackhole background.
Furthermore, one can explore the interplay between the relative signs of $\wp_1$ and  $\wp_2$ such that the two interaction terms can cancel each other either at the boundary or near the horizon. For instance, if we consider a solution with the background parameters $\chi^{(1)}=2$, $T/\mu=0.009$, $k_1/\mu=0.8$,\;$k_2/\mu=0.2$ and $m_{\phi}^2=0$, we can have the effective coupling $(\wp_1 + \wp_2 |\phi|^2)$ to be zero, negative or positive for different $\wp_1$ and $\wp_2$ values. 
We have explored such a scenario in \figurename{ \ref{interplayP2}} where the plots show the cases when $(\wp_1 + \wp_2 |\phi|^2)\lessgtr 0$ going through the point $(\wp_1 + \wp_2 |\phi|^2)=0$. For fixed $\wp_2=-0.1$ and varying $\wp_1$ such that the effective coupling  becomes large positive value of  we saw a gap spectrum similar to \figurename{ \ref{densityp1}}(a) and for large negative values, we observed the presence of  secondary FS with the outer surface showing almost of negligible density.
 
\section{Effect of mass on the fermionic spectral function }
\begin{figure}[htbp]
	\centering
	\includegraphics[width=\columnwidth]{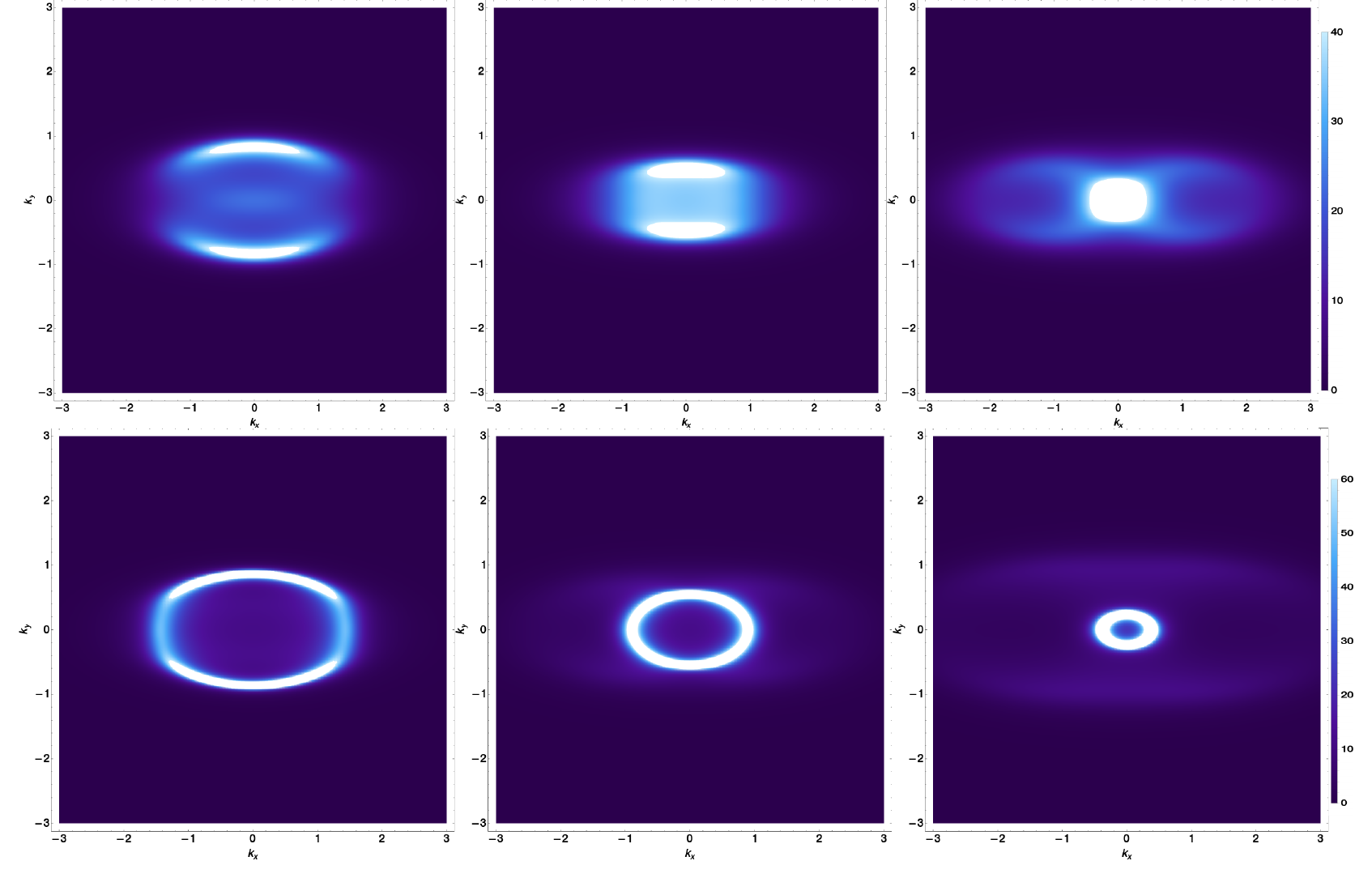}
	\caption{Spectral density with background parameters  $T/\mu=0.009$, $k_1/\mu=0.8$,\;$k_2/\mu=0$, $m_{\phi}^2=0$ and, $\chi^{(1)}=2$. Here, the fermion mass $m_{\psi}=1/4\;,\text{and charge}\;q=1$. In the top panel we vary $\wp_2=-0.1,-0.2,-0.3$ (left to right) and fixing $\wp_1=0.5$. In the bottom $\wp_1=0$ and $\wp_2=-0.1,-0.2,-0.3$ from left to right.}
	\label{nonzeromassvaryp3andp1p3}
\end{figure}

For completeness, we also checked the case where fermion mass is non zero with $m_{\psi}=1/4$ within the mass window $m_{\psi}<1/2$, where we have the freedom to choose either the coefficient $A_j$ or $D_j$ as the sources for fermionic operators. We did not find any significant differences and interesting effects on the spectral function other than what we observed for massless fermions. Nonetheless we present some of the results in \figurename{ \ref{nonzeromassvaryp3andp1p3}}. \par

On the other hand, when the scalar mass is within the BF bound $-9/4\le m_{\phi}^2<0$, for mass $m_{\phi}^2$ close to $-9/4$, for example, $m_{\phi}^2=-2$, the contribution from our coupling to the boundary fermionic spectral function was highly suppressed, however, when $m_{\phi}$ approaches zero, the effect is much more significant as we have seen from the results for $m_{\phi}^2=0$. We checked for $m_{\phi}^2=-11/16$, which correspond to $\alpha_-=1/4$ and $\alpha_+=11/4$, and we obtained similar results.

\section{Conclusion}
We have considered the anisotropic lattice background, namely the Q-lattice, and explored  the fermionic spectral function in detail by considering two specific fermion-gauge field couplings. In addition to the well-studied dipole coupling, we have also introduced the coupling containing the background scalar field  $|\phi|^2$, which gives the interesting features.
 
In this paper, we adopt a similar generalisation considered in our previous constructions by considering both types of coupling into the fermionic spectral function. We further introduce the scalar field interaction with the fermion and introduce the discrete spatial translation symmetry in two dimensions on the boundary field theory. Apart from producing the anisotropic background, the non-normalizable mode of the scalar field plays the role of controlling the dipole coupling parameter in the bulk. We observed some interesting features on the spectral function due to $(\wp_1 + \wp_2 |\phi|^2)$ coupling. For the first case, the background is anisotropic without any Lorentz violation, and the spectral function passes through some interesting phases (see Fig.\ref{figure8omk}), where phenomena like spectral transfer, Fermi surface smearing happen when simultaneously turning on both the couplings $(\wp_1,\wp_2)$.

\vskip 3mm
\noindent
{\bf Acknowledgements}
\vskip 3mm
\noindent
We are very much grateful to our colleague Dr U. N. Maiti, for illuminating discussions on the experimental aspects of our present work.

 \bibliographystyle{elsarticle-num}
\bibliography{letter.bib}

\begin{thebibliography}{10}
\expandafter\ifx\csname url\endcsname\relax
  \def\url#1{\texttt{#1}}\fi
\expandafter\ifx\csname urlprefix\endcsname\relax\def\urlprefix{URL }\fi
\expandafter\ifx\csname href\endcsname\relax
  \def\href#1#2{#2} \def\path#1{#1}\fi

\bibitem{Maldacena:1997re} J.~ M.~ Maldacena, \emph{The Large $N$ limit of superconformal field theories and supergravity}, \emph{Int. J. Theor. Phys.} {\bf 38} (1999) 1113 [{\emph{Adv. Theor. Math. Phys.}}{\bf 2} (1998) 231].

\bibitem{Witten:1998qj}  E.~Witten, \emph{Anti-de Sitter space and holography},
\emph{Adv. Theor. Math. Phys.}  {\bf 2} (1998) 253,

\bibitem{Bhattacharyya:2008jc} 
S.~Bhattacharyya, V.~E.~Hubeny, S.~Minwalla and M.~Rangamani,
\emph{Nonlinear Fluid Dynamics from Gravity},
\emph{JHEP} {\bf 02} (2008) 045.



\bibitem{Faulkner:2009wj} 
T.~Faulkner, H.~Liu, J.~McGreevy and D.~Vegh, \emph{Emergent quantum criticality, Fermi surfaces, and AdS(2)},
\emph{Phys. Rev. D.}  {\bf 83}  (2011) 125002.

\bibitem{Cubrovic:2009ye} 
M.~Cubrovic, J.~Zaanen and K.~Schalm, \emph{String Theory, Quantum Phase Transitions and the Emergent Fermi-Liquid},
\emph{Science} {\bf 325} (2009) 439.


\bibitem{Herzog:2007ij} 
C.~P.~Herzog, P.~Kovtun, S.~Sachdev and D.~T.~Son,
\emph{Quantum critical transport, duality, and M-theory},
\emph{Phys. Rev. D.} {\bf 75} (2007) 085020.


\bibitem{Hartnoll:2008vx} 
S.~A.~Hartnoll, C.~P.~Herzog and G.~T.~Horowitz,
\emph{Building a Holographic Superconductor}, \emph{Phys. Rev. Lett.}  {\bf 101} (2008) 031601.

\bibitem{Hartnoll:2008kx} 
S.~A.~Hartnoll, C.~P.~Herzog and G.~T.~Horowitz,
\emph{Holographic Superconductors},
\emph{JHEP} {\bf 12} (2008) 015,

\bibitem{Horowitz:2010gk} 
G.~T.~Horowitz,
\emph{Introduction to Holographic Superconductors}, \emph{Lect. Notes Phys.}  {\bf 828} (2011) 313.



\bibitem{Liu:2009dm} 
H.~Liu, J.~McGreevy and D.~Vegh,
\emph{Non-Fermi liquids from holography},
\emph{Phys. Rev. D} {\bf 83} (2011) 065029.

\bibitem{Lee:2008xf} 
S.~S.~Lee,
\emph{A Non-Fermi Liquid from a Charged Black Hole: A Critical Fermi Ball},
\emph{Phys. Rev. D} {\bf 79} (2009) 086006.

\bibitem{Faulkner:2009am} 
T.~Faulkner, G.~T.~Horowitz, J.~McGreevy, M.~M.~Roberts and D.~Vegh, \emph{Photoemission `experiments' on holographic superconductors},
\emph{JHEP} {\bf 03} (2010)  121.

\bibitem{Guarrera:2011my} 
D.~Guarrera and J.~McGreevy, \emph{Holographic Fermi surfaces and bulk dipole couplings}, arXiv:1102.3908 [hep-th].

\bibitem{Faulkner:2011tm} 
T.~Faulkner, N.~Iqbal, H.~Liu, J.~McGreevy and D.~Vegh, \emph{Holographic non-Fermi liquid fixed points}, \emph{Phil. Trans. Roy. Soc. A}  {\bf  369}  (2011) 1640.

\bibitem{Vanacore:2015poa} G.~Vanacore, S.~T.~Ramamurthy and P.~W.~Phillips,
\emph{Evolution of Holographic Fermi Arcs from a Mott Insulator}, \emph{JHEP} {\bf 1809} (2018) 009.


\bibitem{Edalati:2010ww} 
M.~Edalati, R.~G.~Leigh and P.~W.~Phillips, \emph{Dynamically Generated Mott Gap from Holography},
\emph{Phys. Rev. Lett.}  {\bf 106} (2011) 091602.
\bibitem{PhysRevD.90.126013} J.~ Alsup, E.~ Papantonopoulos, G.~ Siopsis, K.~ Yeter, \emph{Duality between zeroes and poles in holographic systems with massless Fermions and a dipole coupling},\emph{Phys. Rev. D} {\bf 90} (2014) 126013. 



\bibitem{Damascelli:2003bi} 
A.~Damascelli, Z.~Hussain and Z.-X.~Shen, \emph{Angle-resolved photoemission studies of the cuprate superconductors},
\emph{Rev. Mod. Phys.}  {\bf 75} (2003)  473.

\bibitem{2011PhRvL.106l7005K} P. D. C. King, J. A. Rosen, W. Meevasana, A. Tamai, E. Rozbicki, R. Comin, G. Levy, D. Fournier, Y. Yoshida, H. Eisaki, K. M. Shen, N. J. C. Ingle, A. Damascelli and F. Baumberger, \emph{Structural origin of apparent Fermi surface pockets in angle-resolved photoemission of Bi$_2$Sr$_{{\rm{2-x}}}$La$_{{\rm{x}}}$CuO$_{{\rm{6+\delta}}}$}, \emph{Phys. Rev. Lett.} {\bf 106} (2011) 127005. 

\bibitem{bednorz} J. G. Bednorz and K. A. Muller, Possible high $T_c$ superconductivity in the Ba-La-Cu-O system, Z. Phys. B 64 (1986) 189.


\bibitem{1998Natur.392..157N} 
M. R. Norman, H. Ding, M. Randeria, J. C. Campuzano, T. Yokoya, T. Takeuchi, T. Takahashi, T. Mochiku, K. Kadowaki, P. Guptasarma and D. G. Hinks, \emph{Destruction of the Fermi surface in underdoped high-$T_c$ superconductors}, \emph{Nature}{\bf 392} (1998) 157.

\bibitem{2006PhRvB..74v4510Y} T. Yoshida, X. J. Zhou, K. Tanaka, W. L. Yang, Z. Hus- sain, Z. X. Shen, A. Fujimori, S. Sahrakorpi, M. Lin- droos, R. S. Markiewicz, A. Bansil, S. Komiya, Y. Ando, H. Eisaki, T. Kakeshita and S. Uchida, \emph{Doping Evolution of the Underlying Fermi Surface in La$_{\rm{2-x}}$Sr$_{\rm{x}}$CuO$_4$}, \emph{Phys. Rev. B} {\bf{74}} (2006) 224510. 

\bibitem{Cremonini:2018xgj} 
S.~Cremonini, L.~Li and J.~Ren, \emph{Holographic Fermions in Striped Phases,}
\emph{JHEP} {\bf 12} (2018) 080.

\bibitem{Seo:2018hrc}
Y.~Seo, G.~Song, Y.~H.~Qi and S.~J.~Sin, \emph{Mott transition with Holographic Spectral function,}
\emph{JHEP} {\bf 08} (2018) 077.

\bibitem{PhysRevB.99.161113} Y. Wu, N. H. Jo, L-L Wang, C. A. Schmidt, K. M. Neilson, B. Schrunk, P. Swatek, A. Eaton, S. L. Bud'ko, P. C. Canfield  and A. Kaminski, \emph{Fragility of Fermi arcs in Dirac semimetals}, \emph{Phys. Rev. B} {\bf{99}} (2019).	


\bibitem{YangL}	Yang, L., Liu, Z., Sun, Y. et al. \emph{Weyl semimetal phase in the non-centrosymmetric compound TaAs}. {\it Nature Phys} {\bf 11}, 728-732 (2015). 

\bibitem{brillaux2020Fermi}
Eric Brillaux and Andrei A. Fedorenko,
\emph{Fermi arcs and surface criticality in dirty Dirac materials}, arXiv:2009.12138 [cond-mat.mes-hall].

\bibitem{Su-Yang} Su-Yang Xu et al., \emph{Discovery of a Weyl Fermion semimetal and topological Fermi arcs
	,} \emph{Science}{\bf349}, (2015) 613.

\bibitem{Mehdi} M. Kargarian, M. Randeria and Y-Mi Lu, \emph{Are the surface Fermi arcs in Dirac semimetals topologically protected?
	,} \emph{PNAS}{\bf 113} (2016) 8648.

\bibitem{MZHazan} M. Z. Hasan and C. L. Kane, \emph{Colloquium: Topological insulators}, \emph{Rev. Mod. Phys.}{\bf 82} (2010) 3045.


\bibitem{Binghai} B. Yan and C. Felser, \emph{Topological Materials: Weyl Semimetals,} \emph{Ann. Rev. Cond. Matt. Phys.}{\bf 8} (2017) 337.

\bibitem{PhysRevB.83.205101} X. Wan, A. M. Turner, A. Vishwanath, and S. Y. Savrasov, \emph{Topological semimetal and Fermi-arc surface states in the electronic structure of pyrochlore iridates,}\emph{Phys. Rev. B} {\bf 83} (2011) 205101. 

\bibitem{BaiqingLv} B. Lv, T. Qian and H. Ding, \emph{Angle-resolved photoemission spectroscopy and its application to topological materials,} \emph{Nat Rev Phys}{\bf 1} (2019) 609-626.




\bibitem{PhysRevB.76.174501} M. R. Norman, A. Kanigel, M. Randeria, U. Chatterjee and J. C. Campuzano, \emph{Modeling the Fermi arc in underdoped cuprates}, \emph{Phys. Rev. B} {\bf 76} (2007) 174501. 

\bibitem{PhysRevB.73.174501} K.-Y. Yang, T. M. Rice and F.-C. Zhang, \emph{Phenomenological theory of the pseudogap state}, \emph{Phys. Rev. B} {\bf 73}  (2006) 174501. 

\bibitem{PhysRevB.86.115118} S. Hong and P. Phillips, \emph{Towards the standard model for Fermi arcs from a Wilsonian reduction of the Hubbard model}, \emph{Phys. Rev. B} {\bf 86} (2012) 115118. 

\bibitem{PhysRevB.74.125110} T. D. Stanescu and G. Kotliar, \emph{Fermi arcs and hidden zeros of the Green function in the pseudogap state}, \emph{Phys. Rev. B} {\bf 74} (2006) 125110.	


\bibitem{Hartnoll:2009sz} 
S.~A.~Hartnoll,
\emph{Lectures on holographic methods for condensed matter physics},
\emph{Class. Quant. Grav.}  {\bf 26} (2009) 224002,

\bibitem{Herzog:2009xv} 
C.~P.~Herzog,
\emph{Lectures on Holographic Superfluidity and Superconductivity},
\emph{J. Phys. A} {\bf 42} (2009) 343001,

\bibitem{McGreevy:2009xe} 
J.~McGreevy,
\emph{Holographic duality with a view toward many-body physics},
\emph{Adv. High Energy Phys.}  {\bf 2010} (2010) 723105,

\bibitem{Zaanen:2015oix} J.~ Zaanen, Y-W~ Sun,  and Y.~Liu and K.~Schalm, \emph{Holographic duality in Condensed Matter Physics}; \emph{Cambridge Univ. Press} (2015)

\bibitem{Hartnoll:2016apf} S.~A.~Hartnoll, A.~Lucas and S.~ Sachdev, \emph{Holographic Quantum Matter}; \emph{MIT Press} (2018).



\bibitem{Vegh:2010fc} 
D.~Vegh, \emph{Fermi arcs from holography}, arXiv:1007.0246 [hep-th].

\bibitem{Benini:2010qc} 
F.~Benini, C.~P.~Herzog and A.~Yarom, \emph{Holographic Fermi arcs and a $d$-wave gap}, \emph{Phys. Lett. B} {\bf 701} (2011) 626. 


\bibitem{Chakrabarti:2019gow}
S.~Chakrabarti, D.~Maity and W.~Wahlang,
\emph{Probing the Holographic Fermi Arc with scalar field: Numerical and analytical study},
\emph{JHEP}{\bf{07}}  (2019) 037.


\bibitem{horo1} G. T. Horowitz, J. E. Santos, and D. Tong, \emph{Optical Conductivity with
	Holographic Lattices}, \emph{JHEP} \textbf{07} (2012) 168. 
\bibitem{horo2} G. T. Horowitz, J. E. Santos, and D. Tong, \emph{Further Evidence for
	Lattice-Induced Scaling}, \emph{JHEP} \textbf{11} (2012) 102. 
\bibitem{horo3} G. T. Horowitz and J. E. Santos, \emph{General Relativity and the Cuprates}, \emph{JHEP}{\bf 06} (2013) 087.

\bibitem{donos4} A. Donos and S. A. Hartnoll, \emph{Interaction-driven localization in holography},
\emph{Nature Phys.} \textbf{9} (2013) 649-655.


\bibitem{Donos:2013eha}
A.~Donos and J.~P.~Gauntlett,
\emph{Holographic Q-lattices},
JHEP \textbf{04}, 040 (2014).



\bibitem{Ling2014} 
Y. ~Ling , P. ~Liu , C. ~Niu ,J-P.~ Wu and, Z-Y. ~Xian, \emph{Holographic Fermionic system with dipole coupling on Q-lattice},
\emph{JHEP}{\bf 12}  (2014) 149.	
%

\bibitem{SGubser:2020}
S.~S.~Gubser, J.~Ren, \emph{Analytic fermionic Green’s functions from holography
,}
\emph{PhysRevD} \textbf{86} (2012).

\bibitem{Iliasov:2019pav}
A.~Iliasov, A.~A.~Bagrov, M.~I.~Katsnelson and A.~Krikun, \emph{Anisotropic destruction of the Fermi surface in inhomogeneous holographic lattices,}
\emph{JHEP} \textbf{01} (2020) 065.

\bibitem{Hyun-Sik:2020}
H.~Jeong, K.~Y.~Kim, Y.~Seo,S.~J.~Sin and S.~Wu, \emph{Holographic spectral functions with momentum relaxation
,}
\emph{PhysRevD} \textbf{102} (2020).

\bibitem{Cremonini2019} 
Sera Cremonini, Li Li, Jie Ren, \emph{Spectral Weight Suppression and Fermi Arc-like Features with Strong Holographic Lattices},
\emph{JHEP}{ \bf 09}  (2019) 014.

\bibitem{Balm2020} 
F.~Balm, et.al., \emph{Isolated zeros destroy Fermi surface in holographic models with a lattice},
\emph{JHEP}{ \bf 01}  (2020) 151.


\bibitem{Trefethen}
L. N.~Trefethen, 
\emph{Spectral methods in MATLAB}, \emph{SIAM}, Philadelphia, (2000). 

\bibitem{Boyd}
J. P. ~Boyd, \emph{Chebyshev and Fourier spectral methods,} Lecture Notes in Engineering, {\emph Springer} (1989). 

\bibitem{Andrade:2017jmt}
T.~Andrade,
\emph{Holographic Lattices and Numerical Techniques}, arXiv:1712.00548 [hep-th].

\bibitem{Krikun:2018ufr}
A.~Krikun, \emph{Numerical Solution of the Boundary Value Problems for Partial Differential Equations. Crash course for holographer},


\bibitem{Eoh} E. Oh, Y. Seo, T. Yuk and S-J. Sin, \emph{Ginzberg-Landau-Wilson theory for flat band, Fermi-arc and surface states of strongly correlated systems}. \emph{JHEP}{\bf{01}} (2021) 053. 

\bibitem{DMarchenko} D. Marchenko, D. V. Evtushinsky, E. Golias, A. Varykhalov, Th. Seyller, and O. Rader, \emph{Extremely flat band in bilayer graphene,} \emph{Sci. Adv.}{\bf 4} (2018) 11.

\bibitem{PhysRevB.103.024529} K. ~Yang, \emph{Exactly solvable model of Fermi arcs and pseudogap,} \emph{Phys. Rev. B}{ \bf 2} (2021).


\bibitem{transfer} A. Chikamatsu {\it et al.},  \emph{Gradual disappearance of the Fermi surface near the metal-insulator transition in 
	$\rm{La}_{1-x}\rm{Sr}_{x}\rm{MnO}_{3}$ thin films}, \emph{Phys. Rev.}{\bf B76}, 201103 (2007).



\bibitem{FSsmearing} H. C. Robarts {\em et al.}, \emph{Extreme Fermi Surface Smearing in a Maximally Disordered Concentrated Solid Solution,} \emph{Phys. Rev. Lett.}{\bf  124} (2020) 046402. 	

\end{thebibliography}

\end{document}